\documentclass[12pt]{article}
\usepackage{times}
\usepackage{geometry}
\usepackage[utf8]{inputenc}
\usepackage{enumitem,amssymb}
\usepackage{ragged2e}
\usepackage{xspace}
\newlist{thematic}{itemize}{8}
\setlist[thematic]{label=$\square$}
\usepackage{pifont}
\usepackage[backend=biber,style=numeric-comp,sorting=none]{biblatex}
\usepackage{hyperref}
\usepackage{graphicx}
\usepackage{floatrow}
\usepackage{caption}

\usepackage{sectsty}

\sectionfont{\fontsize{16}{15}\selectfont}

\DeclareCaptionFormat{myformat}{#1#2#3\hrulefill}

\def\justifying{%
  \rightskip=0pt
  \spaceskip=0pt
  \xspaceskip=0pt
  \relax
}

\begin{document}

\captionsetup[figure]{labelfont={bf},name={Figure},labelsep=period}

\raggedright
Heliosphysics 2024 Science White Paper \linebreak
\Large
\linebreak
\textbf{Heliosphere Meets Interstellar Medium, in a Galactic Context} \linebreak
\normalsize

\noindent \textbf{Category:} Basic Research \\ 
\indent \textbf{Primary Topic:} Emerging Opportunities (Interstellar Medium and Astrospheres) \\
\indent \textbf{Secondary Topic:} Outer Heliosphere \linebreak


\textbf{Principal Author:}

Name: Stella Koch Ocker	
 \linebreak						
Institution: Cornell University
 \linebreak
Email: sko36@cornell.edu \vspace{0.1in}

\textbf{Co-authors:} James Cordes$^{(1)}$, Shami Chatterjee$^{(1)}$, Jeffrey Hazboun$^{(2,3)}$, Timothy Dolch$^{(4,5)}$, Daniel Stinebring$^{(6)}$, Dustin Madison$^{(7)}$, Stephen White$^{(8)}$, Gregory Taylor$^{(9)}$, Natalia Lewandowska$^{(10)}$, Michael Lam$^{(11)}$
  \linebreak
$^{(1)}$\textit{Cornell University}, $^{(2)}$\textit{Oregon State University},
$^{(3)}$\textit{University of Washington Bothell},
$^{(4)}$\textit{Hillsdale College},
$^{(5)}$\textit{Eureka Scientific Inc.},
$^{(6)}$\textit{Oberlin College}, 
$^{(7)}$\textit{University of the Pacific}, 
$^{(8)}$\textit{Air Force Research Lab},
$^{(9)}$\textit{University of New Mexico},
$^{(10)}$\textit{State University of New York Oswego},
$^{(11)}$\textit{Rochester Institute of Technology}
\linebreak

\justifying

\textbf{Synopsis:} The physical conditions within our heliosphere are driven by the Sun’s motion through an evolving interstellar environment that remains largely unexplored. The next generation of outer heliosphere and interstellar explorers will answer fundamental questions about the heliosphere’s relationship with the very local interstellar medium (VLISM) by diving deeper into the Sun’s interstellar surroundings. The impact of these future missions will be vastly enhanced by concurrent, interdisciplinary studies that examine the direct connections between conditions within the heliosphere, the heliosphere’s immediate interstellar environment, and the larger-scale Galactic ISM. Comparisons of the heliosphere and VLISM to their analogs across the Galaxy will constrain the global processes shaping both stellar astrospheres and their sustained impact on the ISM. 

\begin{figure}[hb!]
    \centering
    \includegraphics[width=0.9\textwidth]{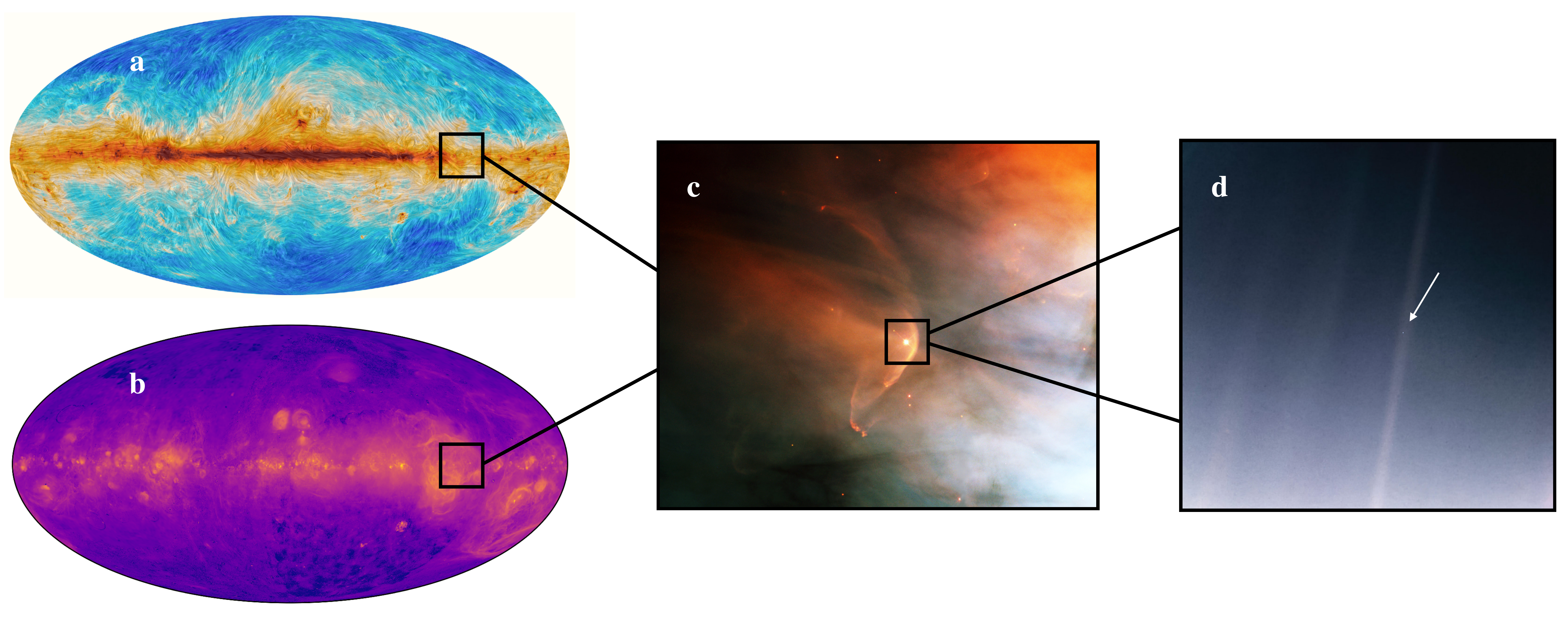}
    \caption{a) Galactic dust polarization map \cite{planck_dustmap}. b) The Galaxy in H$\alpha$ \cite{finkbeiner2003}. c) Bow shock of the star LL Ori (credit: NASA/ESA). d) Last image of Earth from Voyager 1 (credit: NASA/JPL).}
    \label{fig:cover_image}
\end{figure}

\thispagestyle{empty}

\pagebreak

\setcounter{page}{1}

\section{Introduction}

The mid-2012 arrival of Voyager 1 in the VLISM \cite{2013Sci...341.1489G}, the region of interstellar space just outside of the heliosphere, constituted the beginning of a new era of heliophysics research. In just 10 years, Voyager 1 (V1), and later Voyager 2 (V2), have demonstrated that the heliopause is a highly non-uniform, porous boundary between the heliosphere and VLISM \cite{2013Sci...341..147B,2013Sci...341..150S,2019NatAs...3.1007B,2019NatAs...3.1013S}. 
Independent astronomical observations of the local interstellar cloud (LIC) and the Local Bubble (LB) \cite{2019ApJ...886...41L,2022SSRv..218...16L}, combined with magnetohydrodynamic (MHD) modeling of the heliosphere \cite{2022SSRv..218...36K}, have begun to contextualize Voyager's findings in terms of the long-term journey and evolution of the heliosphere's global shape, size, and energetics. Despite this progress, critical questions about the heliosphere's relationship with the ISM remain unanswered, and even exacerbated. In the next decade, a combination of outer heliosphere and VLISM missions with astronomical observations of the local ISM and other stars will answer fundamental, interconnected questions:
\begin{itemize}[itemsep=0.5pt] \justifying
    \item What is the nature and extent of the Sun’s interaction with the ISM, and how does this relationship compare to other stars?
    \item Can a spacecraft explorer travel beyond the VLISM, and if so, what will it reveal about the content and energetics of the Galactic ISM?
    \item How unusual or typical is the Sun’s interstellar environment, compared to that of other astrospheres throughout the Galaxy?
\end{itemize} 

Addressing these questions requires the assessment of both stochastic (turbulent) and deterministic processes near stellar-ISM interfaces, in comparison to analogous processes in the diffuse, larger-scale ISM. Novel interdisciplinary approaches that connect spacecraft data to remote observations of other stars and the ISM will illuminate not only the processes driving the global structure of the heliosphere, but also the extent to which these processes are common features of stellar systems across the Galaxy. \vspace{0.1in}

We focus on two key themes: spacecraft exploration of the VLISM, and complementary studies of the heliosphere, local ISM, and other star-ISM interactions enabled by observations of high-velocity stars and compact radio sources, including pulsars and extragalactic quasars that act as beacons through all ionized components of the ISM.


\section{The Dynamic Heliosphere-ISM Interaction}

\begin{figure}
    \centering
    \includegraphics[width=0.8\textwidth]{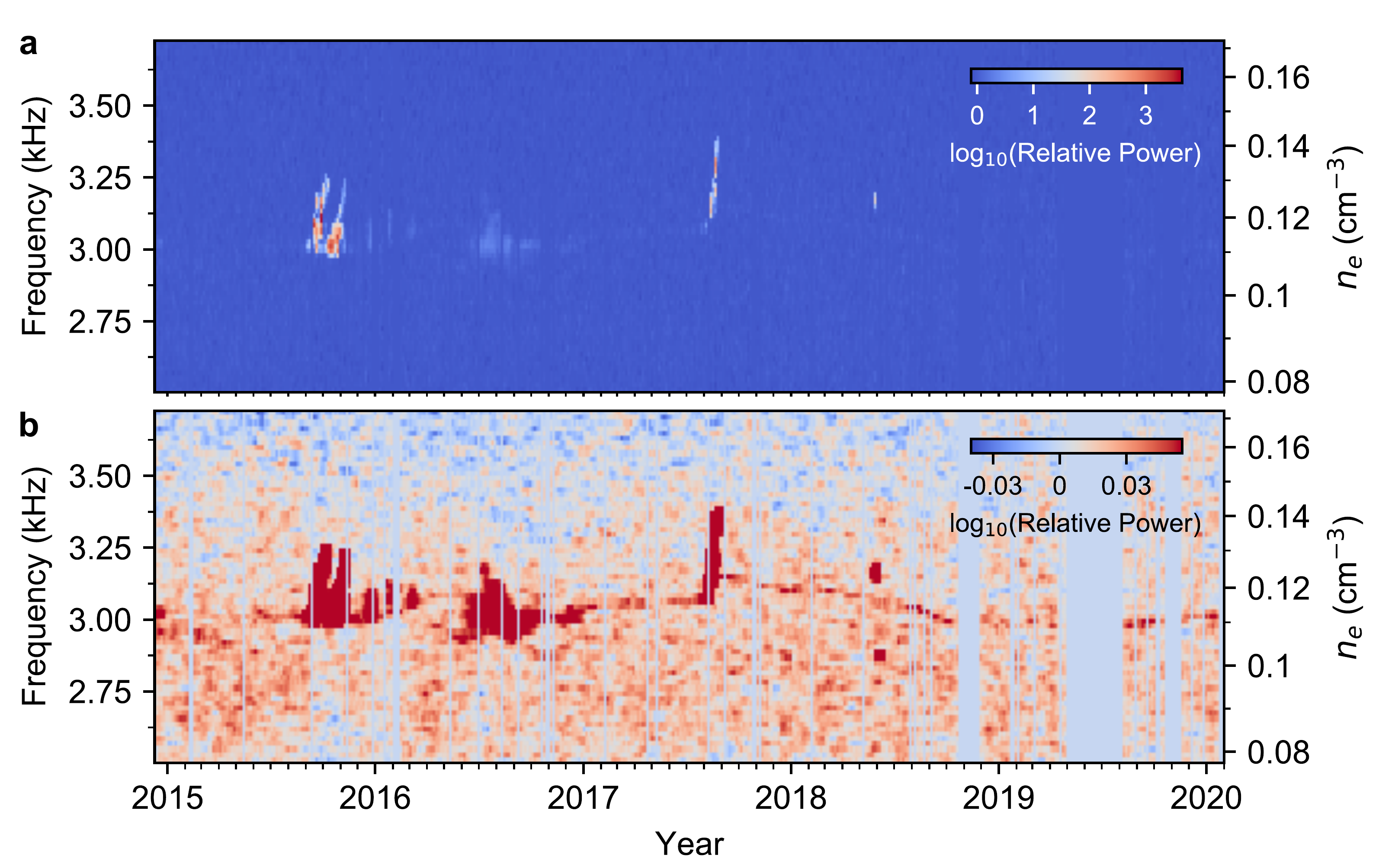}
    \captionsetup{format=myformat}
    \caption{a) VLISM electric field intensity in frequency (kHz) vs. time (years) measured by the Voyager 1 Plasma Wave Subsystem. Bright patches of power are solar shock-triggered plasma oscillation events. b) Same as a) but smoothed in time and shown with a restricted color stretch. The narrow line between the solar-origin plasma oscillation events corresponds to the persistent plasma wave emission that enables a near-continuous sampling of the plasma density. This figure was adapted from Ref.\cite{2021NatAs.tmp...84O}. \textit{Interstellar Probe} will be capable of resolving this narrow plasma line with much higher signal-to-noise and over a greater extent of the VLISM.}
    \label{fig:pws_detection}
\end{figure}

\textbf{Necessity of Interstellar Spacecraft.} The \textit{Voyager Interstellar Mission} (\textit{VIM}) is currently the only source of direct measurements in the VLISM. These \textit{in situ} data are incomparable to any other indirect probe of the outer heliosphere and VLISM, and have led to key discoveries about their relationship, including:
\begin{itemize}[itemsep=0.5pt] \justifying
    \item The magnetic boundary of the heliopause is not uniform. While both V1 and V2 traversed comparable plasma density discontinuities at the heliopause \cite{2019NatAs...3.1024G}, V1 saw dramatic magnetic field fluctuations \cite{2013Sci...341..147B} that were not later observed by V2 \cite{2019NatAs...3.1007B}. V2 also observed a cosmic ray boundary layer outside the heliopause, which was not seen by V1 \cite{2019NatAs...3.1013S}. Immediately outside the heliopause, V2 found VLISM plasma about $2\times$ hotter than expected ($\sim 40,000$ K) \cite{2019NatAs...3.1019R}. 
    \item Shocks of solar origin propagate beyond the heliopause, generating plasma oscillation events, magnetic field jumps, and cosmic ray bursts \cite{2021AJ....161...11G}. While these shocks were observed regularly (about once a year) by V1 from the heliopause at 121 au out to 145 au \cite{2021AJ....161...11G}, no such events have been observed since. It is unclear whether the recent absence of shocks is related to the solar cycle or whether it signifies the extent of solar influence on the VLISM.
    \item The unexpected detection of faint, persistent plasma wave emission by V1 has enabled a highly sampled density map of the VLISM \cite{2021NatAs.tmp...84O}, and demonstrates that the VLISM electron distribution has a supra-thermal tail \cite{2021ApJ...921...62G,2022A&A...658L..12M}; see Figure~\ref{fig:pws_detection}.
    \item The V1 observation of a large magnetic pressure front and density jump at about 150 au \cite{2021ApJ...911...61B} without any accompanying plasma oscillations raises the question of whether V1 has identified another boundary layer in the VLISM.
\end{itemize}

These discoveries have raised questions with significant implications for conditions within the heliosphere: What drives the non-uniformity of the outer heliosphere's structure? How far does the VLISM extend? Does the heliosphere have a bow shock (and if so, what is its nature)? Do shock waves from other stars perturb the VLISM?\vspace{0.1in} 

Direct spacecraft measurements are critical to answering these questions. While \textit{VIM} is expected to continue operating for the next few years, the launch of a new \textit{Interstellar Probe} mission \cite{2022AcAau.196...13M,BRANDT2022} in the next decade would be pivotal for heliophysics and VLISM research. \vspace{0.1in} 

\textit{Interstellar Probe} will provide the first high-resolution map of plasma densities, magnetic fields, and energetic particles spanning both the inner and outer heliosphere and the VLISM (out to at least 300 au) \cite{2022AcAau.196...13M,BRANDT2022}. Anticipated results include: The effect of solar cycle on the size and structure of the heliosphere and its boundary with the VLISM; signatures of interstellar-origin shocks and structures; the evolution of turbulence and its degree of anisotropy from outer heliosphere to VLISM; and constraints on the heliospheric bow shock's existence. By exploring a novel trajectory through the heliosphere, \textit{Interstellar Probe} will empirically test MHD models for the global structure of the heliosphere and its evolution as a function of interstellar environment.\footnote{See the \textit{Interstellar Probe} white paper (lead author Pontus Brandt) for more detail about the mission concept.} \vspace{0.1in}

\begin{figure}
    \centering
    \includegraphics[width=\textwidth]{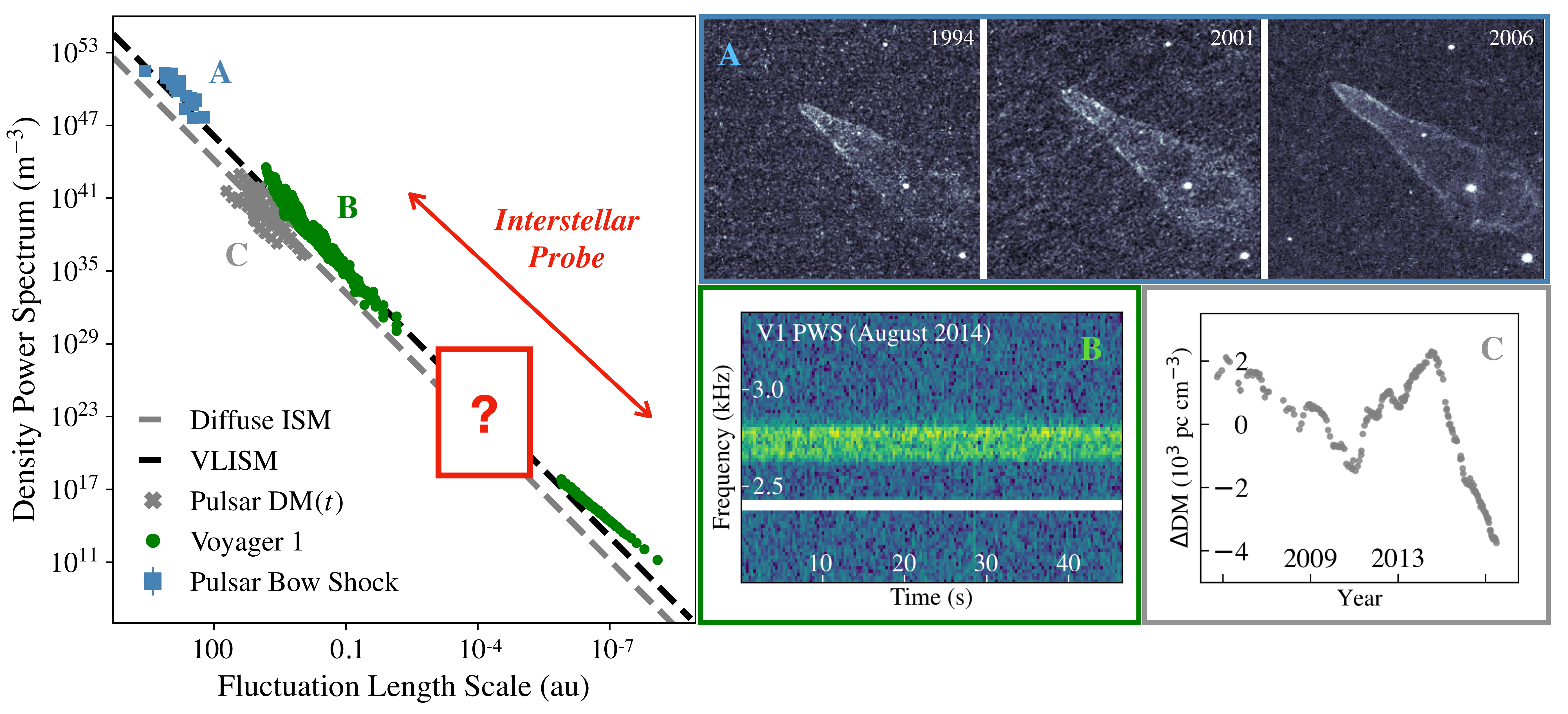}
    \captionsetup{format=myformat}
    \caption{The lefthand plot shows the composite power spectrum of turbulent, interstellar electron density fluctuations as a function of length scale (au), inferred from three different \textit{in situ} and path-integrated probes illustrated on the right: A. \textit{In situ} H$\alpha$ images of the pulsar bow shock known as the Guitar Nebula; B. Plasma densities measured \textit{in situ} using plasma oscillations detected by the Voyager 1 Plasma Wave Subsystem (PWS); C. Path-integrated pulsar dispersion measure (DM) variations measured by the NANOGrav Collaboration over months to years. The proposed \textit{Interstellar Probe} will sample the as-yet unexplored region indicated by the red question mark.}
    \label{fig:wavespec}
\end{figure}

The scientific impact of \textit{Interstellar Probe} will extend beyond heliophysics to interstellar astrophysics research. Spacecraft measurements of magnetic field and electron density fluctuations are the only direct probes of interstellar turbulence, which is characterized using magnetic field and plasma density power spectra (see Figure~\ref{fig:wavespec}). V1 has thus far provided the only constraints on the interstellar turbulence power spectrum at kinetic and dissipative scales ($\sim$ kilometers), which cannot be accessed with other remote probes of the ISM (Figure~\ref{fig:wavespec}). However, there is a significant gap in the turbulence spectrum probed by \textit{VIM} between length scales of $\sim 10^3$ km and $\sim 0.01$ au. \textit{Interstellar Probe} will span this gap, thereby providing the first self-consistently measured turbulence spectrum across 10 orders of magnitude in length scale.  \vspace{0.1in}



\textbf{The Unique Value of Astronomical Remote Sensing.} Despite their extraordinary resolving power, spacecraft measurements are limited by the singular sightlines and distances they can probe. Remote observations along many lines-of-sight (LOSs) across the sky are thus required to contextualize \textit{in situ} spacecraft measurements, both in and outside the heliosphere. \vspace{0.1in}

Remote sensing observations can be categorized as either path-integrated or \textit{in situ}. Here we focus mainly on radio pulsars\footnote{Optical emission and absorption lines are also critical path-integrated probes of the VLISM and local clouds; see the white paper led by Linsky and Redfield.}, which undergo dispersion in cold plasma. The resulting frequency-dependent time delay in their pulse times-of-arrival depends on the path-integrated electron column density, i.e. the dispersion measure (DM). Another quantity of interest is the Faraday rotation measure (RM) from polarization measurements, which gives information about the path-integrated magnetic field. Together, DMs and RMs for many LOSs are inputs to tomographic maps of the ionized ISM. Pulsars and other compact radio sources are also scattered by intervening plasma, which is another path-integrated effect discussed further below. The chief remote sensing method for \textit{in situ} observation is imaging. Imaging other stellar bow shocks and astrospheres provides \textit{in situ} constraints on interstellar gas and star-ISM interactions that are complementary to spacecraft measurements of the heliosphere and VLISM (see Section~\ref{sec:galactic}). \vspace{0.1in}


\textbf{Inside the Heliosphere: Using Pulsar Dispersion through the Solar Wind.} 
Regular monitoring of a relatively dense pulsar population across the sky will provide 3D constraints on the solar wind density throughout the heliosphere, in a manner independent and complementary to the myriad space missions studying solar weather.
Pulsar DMs undergo annual variations as their LOSs pass through the solar corona over the course of Earth's orbit. These DM variations are sensitive to solar wind features that are more complex than the spherically symmetric $1/r^2$ component \cite{you+2007,you+2012}. 
Recent work \cite{madison+2019,tiburzi+2019,tiburzi+2021, 2022ApJ...929...39H} has shown that pulsar timing arrays (PTAs) can be valuable probes of the solar wind and its temporal behavior. PTAs survey a much wider swath of the Sun's environment than spacecraft by taking electron column density information for 70+ LOSs every month, thereby probing even the outermost reaches of the solar environment. Lower frequency, high-cadence observing campaigns with, e.g., the Canadian Hydrogen Intensity Mapping Experiment (CHIME) \cite{chime/pulsar2021}, the Low Frequency Array (LOFAR) \cite{lofar2011,lofar2013}, and the Long Wavelength Array (LWA) \cite[][]{lwa_pulsar2015} are also continually adding data that is extremely useful for DM variability studies, since these lower frequencies allow for more accurate DM variation measurements \cite{Kumar:2022bts}. Similar, high-cadence monitoring of RMs will provide complementary magnetic field information \cite{ingleby2007}. \vspace{0.1in}

\begin{figure}
    \floatbox[{\capbeside\thisfloatsetup{capbesideposition={right,center},capbesidewidth=5.5cm}}]{figure}[\FBwidth]
    {\captionsetup{format=myformat} \caption{a) Pulsar radio intensity in time vs. frequency, i.e. the dynamic spectrum (from Ref.\cite{cordes2006}). b) The scintillation arc spectrum formed via Fourier transform of the dynamic spectrum. Each parabolic arc corresponds to a distinct scattering screen. c) Locations of scattering screens along the LOS to this pulsar, inferred with scintillometry (adapted from Ref.\cite{mckee2022}). The closest screen is 5 pc from Earth.}
    \label{fig:scint}}
    {\includegraphics[width=0.65\textwidth]{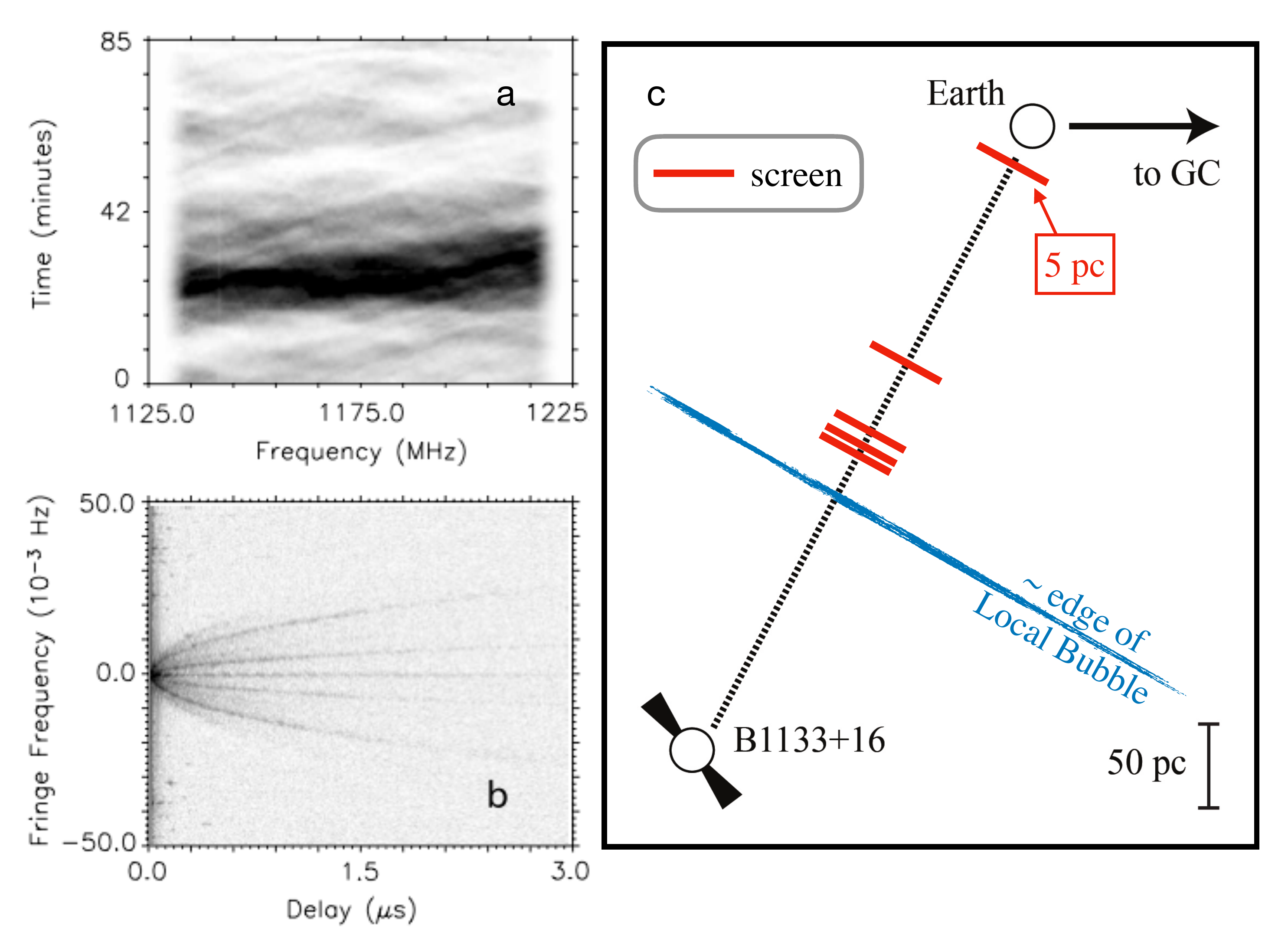}}
\end{figure}

All of these data can add important tests to the well developed solar weather modeling efforts undertaken by the heliophysics community \cite{awsom2014}. These large-scale measurements would allow for long-timescale monitoring of the fast and slow solar wind, giving access to continuous measurements of the solar wind at all solar latitudes. Solar wind models are also a critical element of high-accuracy pulsar astronomy applications, including nanohertz frequency gravitational wave detection (see Astro2020 decadal survey priority area ``New Messengers and New Physics")\footnote{See also the Nanohertz Observatory for Gravitational Waves (NANOGrav) Astro2020 White Paper: \url{https://baas.aas.org/pub/2020n7i195/release/1}} \cite{2022ApJ...929...39H}. \\ \vspace{0.1in} 
\textbf{Outside the Heliosphere: Interstellar Scintillation as Probes of Sub-au Structure in the Local Interstellar Clouds (LICs).} Radio scintillation observations are uniquely sensitive to small-scale structure in the LICs that shape the heliosphere. Radio emission from compact sources like pulsars and extragalactic quasars is scattered by sub-au scale plasma inhomogeneities along their LOSs, leading to constructive and destructive interference between the observed scattered rays. The resulting diffraction pattern changes over time due to the Earth's motion relative to the LOS, a phenomenon known as scintillation. Fourier analyses of pulsar scintillation patterns, a method known as scintillometry, show that scattering appears to occur in confined, discrete regions of plasma along the LOS (see Figure~\ref{fig:scint}), and the precise distances of these scattering regions are constrained by scintillation observations taken months to years apart \cite{mckee2022}.  
The scintillometry method traces processes in ISM plasma from scales of $\sim 1000$ km to $\sim$au, thus representing one of the few remote, path-integrated probes of sub-au scale processes that are otherwise measured by spacecraft. \vspace{0.1in}

A number of pulsars exhibit scintillation from plasma structures within or near the edge of the Local Bubble \cite{stinebring22}, and the presence of multiple scattering regions along some pulsar LOSs suggests these structures have a high number density in the ISM. In one case, the closest scattering region lies just 5 pc away from Earth \cite{mckee2022}, within the LICs whose pressure, temperature, density, and magnetic field determine the heliosphere's current shape and size and the flow of energetic particles past Earth \cite{2022SSRv..218...16L} (see Figure~\ref{fig:scint}). \vspace{0.1in}

Large-scale scintillometry of hundreds of pulsars with new, more capable telescopes (e.g., DSA-2000\footnote{DSA-2000 Astro2020 White Paper: \url{https://baas.aas.org/pub/2020n7i255/release/1}}) will reveal more of these structures, yielding unique information about the formation and sustenance of micro-structure in the local interstellar clouds, including boundary conditions on cloud interfaces, which are poorly understood. Dedicated pulsar timing programs (with CHIME, MeerKAT, DSA-2000, and more) targeting large numbers of nearby pulsars are required to achieve this goal, and are already demonstrated to be highly successful at detecting scintillation from a large fraction of such pulsars when observed \cite{wu2022_lofarscintsurvey, stinebring22}. \vspace{0.1in}

\begin{figure}
    \floatbox[{\capbeside\thisfloatsetup{capbesideposition={right,center},capbesidewidth=5cm}}]{figure}[\FBwidth]
    {\captionsetup{format=myformat} \caption{Lines of sight to pulsars within 500 pc of the Sun with precisely measured parallax distances, mapped on to the Galactic plane (the Galactic Center is to the right). Boxes next to each point list the mean electron density using the DM and parallax distance. Red lines indicate the three objects closer than 200 parsecs.
    \label{fig:spoke}}}
    {\includegraphics[width=0.5\textwidth]{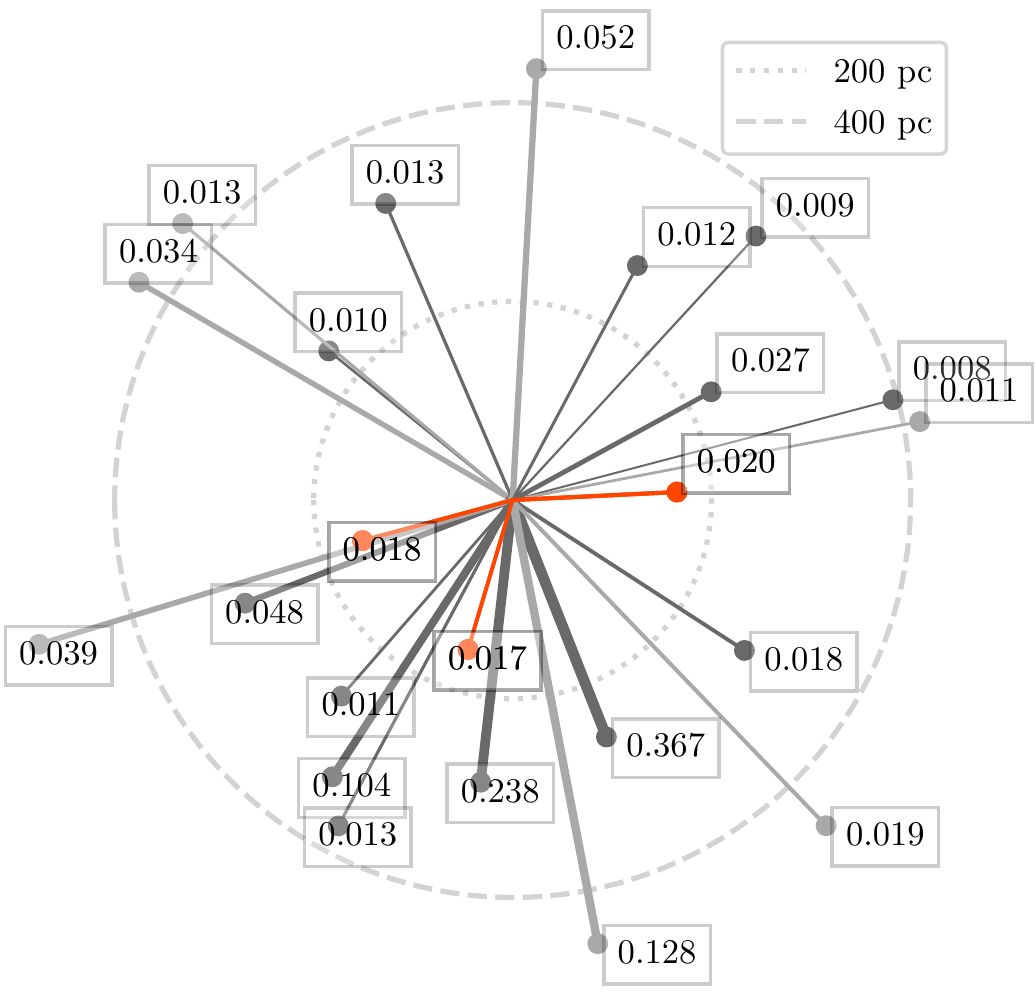}}
\end{figure}

\textbf{From the Heliosphere to the Local Bubble and ISM.} The power of pulsar measurements lies in precision DMs combined with precision parallaxes, which yield an unbiased tracer of the ISM electron column density. Interpretation of DMs does not require knowledge of the gas temperature or species abundances. {\color{black} Telescopes at the lowest frequencies (e.g. the LWA) will produce the highest-precision DMs possible for many pulsars.}
The current sample of pulsar parallaxes within 400 pc of the heliosphere suggests the mean electron density within the LB ranges from 0.009 to 0.018 cm$^{-3}$ (Figure~\ref{fig:spoke}), calling into question the pervasiveness of hot gas within the Bubble \cite{linsky21_stromgren}. \vspace{0.1in}

Hundreds of new pulsar parallaxes (from pulsar timing and very long baseline interferometry) coupled with DMs will yield a high-resolution, tomographic electron density map of the LB and immediate ISM, which is informative about the ionizing radiation field that sustains the LICs. The LB represents just one of many supernova-driven cavities that permeate the Galactic disk. Over the next decade, up to 5,000 pulsars will be discovered, enabling identification and characterization of the overall porosity in the ISM. 

\section{The Galactic Context}\label{sec:galactic}

\textbf{Lessons from Stellar Bow Shocks.} Observations of high-velocity stellar bow shocks present a critical opportunity to constrain in real-time the ISM conditions that also affect the astrospheres of lower velocity stars, including the heliosphere. Bow shocks are observed around stars at a range of life stages, including runaway OB stars and supergiants, and their signatures span radio to X-ray wavelengths \cite{2012A&A...538A.108P, 2015A&A...578A..45P,2012A&A...548A.113D}. Neutron stars provide one test bed for examining the effects of ISM conditions on bow shock evolution, as they are born at speeds ranging from 100s to 1000s of kilometers per second and generally produce bow shocks once they have moved from their supernova remnants into the ISM. One of the few direct methods for detecting neutron star bow shocks is by observing the H$\alpha$ emission produced by collisional excitation of interstellar gas at the bow shock. Figure~\ref{fig:wavespec} shows \textit{Hubble Space Telescope} images (H$\alpha$) of the pulsar bow shock known as the Guitar Nebula \cite{1993Natur.362..133C}. The bow shock morphology evolves over years-long timescales as a direct result of ISM density variations sampled by the rapidly moving pulsar \cite{2002ApJ...575..407C,ocker_bowshocks}. \vspace{0.1in}

Similar ISM variations in plasma density, ionization, and magnetic fields directly impact the flux of cosmic rays and pick-up ions on Earth and the other planets \cite{frisch2011}. As the heliosphere travels through a complex system of interstellar clouds with varying densities, ionization levels, and turbulence, conditions for planetary habitability also evolve. The heliosphere will exit the LIC and enter a new interstellar cloud in the next 2,000 years \cite{2022SSRv..218...16L}, but the precise location of the cloud boundary, and the expected effects on the heliosphere and life within it, remain poorly constrained. The large increase in pulsar (and other) sightlines in the next decade may sharpen our knowledge of that local boundary. \vspace{0.1in}

\textbf{A Galaxy of Astrospheres.} The multi-phase ISM is permeated by billions of stellar astrospheres \cite{wood2004}, most of which host at least one exoplanet. As the known exoplanet population continues to dramatically rise thanks to ongoing searches with new and upcoming telescopes, including the \textit{James Webb Space Telescope}, characterization of their habitability will require consideration of star-ISM interactions analogous to those driving the formation of the heliosphere and the conditions for life within it. ISM conditions vary dramatically across the Galaxy, making continued remote study of the ISM critical to characterization of other stellar astrospheres and their exoplanets.

\section{The Next Decade of Opportunities}

Over the next decade we will begin painting a holistic picture of the heliosphere and its Galactic context, not only via large-scale studies of the solar wind, but also through a global assessment of the heliosphere's interstellar environment and its connection to physical conditions in the broader ISM. This work is only possible through interdisciplinary studies enabled by a combination of spacecraft and ground-based observations, as summarized in the table below.

\vfill

\begin{figure}[hb]
    \centering
    \includegraphics[width=\textwidth]{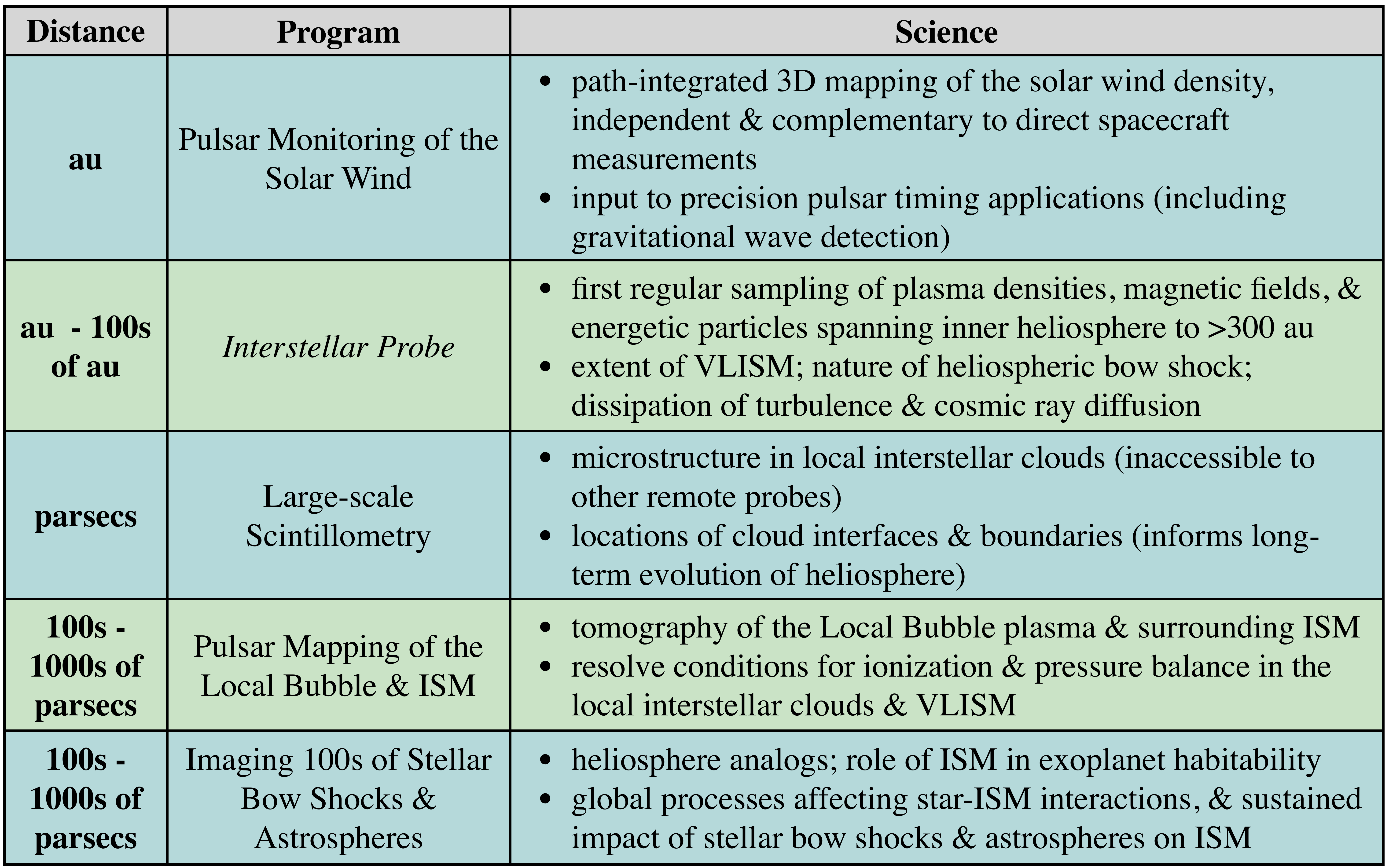}
    \caption*{}
    \label{fig:table}
\end{figure}

\pagebreak

\printbibliography

\end{document}